# Ultrafast Carrier Dynamics in Two-Dimensional Electron Gas-Like K-doped MoS$_2$

*Calley N. Eads,[1] Sara L. Zachritz,[1] Joohyung Park,[1] Anubhab Chakraborty,[1] Dennis L. Nordlund,[2] and Oliver L.A. Monti[1,3]\**

[1] University of Arizona, Department of Chemistry and Biochemistry, 1306 E. University Blvd., Tucson, AZ 85721, USA

[2] SLAC National Accelerator Laboratory, Stanford Synchrotron Radiation Lightsource, 2575 Sand Hill Rd., MS 99, Menlo Park, CA 94025, USA

[3] University of Arizona, Department of Physics, 1118 E. 4th St., Tucson, AZ 85721, USA

**\*Corresponding author:** Oliver L.A. Monti, monti@u.arizona.edu

*Abstract*

Electronic interactions associated with atomic adsorbates on van der Waals layered transition metal dichalcogenides such as $MoS_2$ can induce massive electronic reconstruction that results in the formation of a metallic phase that exhibits characteristics of a two-dimensional electron gas. The impact and mechanism of quantum confinement and reduced dimensionality on the carrier dynamics in such two-dimensional systems is at present not known, but is of paramount importance if they are to find application in optoelectronic devices. Here, we show by a combination of angle-resolved photoemission and advanced x-ray spectroscopies that many-body interactions in reduced dimension drastically shorten carrier lifetimes and reveal how potassium intercalation in $MoS_2$ forces orbital rehybridization to create a two-dimensional electron gas.

*1. Introduction*

Two-dimensional electron gases (2DEGs) form when electrons are confined to quantum well structures and provide an exquisite test-bed to understand electron behavior in reduced dimensions, the underpinning mechanism in modern thin film devices. Historically, 2DEGs have been observed at asymmetric heterojunctions in oxide heterostructures promoting superconductivity and large magnetoresistance[1–4] and semiconductor heterostructures exhibiting enhanced carrier mobilities among others.[5–7] More recently, interest has shifted to atomically-thin layered materials such as graphene,[8,9] black phosphorus[10] and transition metal dichalcogenides (TMDs),[11,12] where electrons are naturally confined to their respective layers. TMDs have received significant attention with not only hosting 2DEG-like states but with the introduction of new physics, including spin-valley coupling[13–15] and topological phases,[16,17] relevant to spintronic and valleytronic devices[18–20] and quantum information processing.[21] The confinement of the wavefunction to individual layers which is central to many of the unique properties and applications of TMDs also makes them ideally suited to potentially support 2DEGs in bulk crystals, as has been suggested in $WSe_2$,[22] $HfSe_2$[23] and $MoSe_2$.[11]

TMDs are composed of $MX_2$ layers (M: metal, X: chalcogen) that stack upon each other to create quasi-2D van der Waals crystals. The TMD structure and bonding scheme result in a distinct electronic structure with layer-dependent electronic properties, including thickness-dependent bandstructure[24,25] and an indirect-to-direct bandgap transition.[26,27] Modification of TMDs by doping and surface deposition has proved an effective means of tailoring their properties. Adsorption of alkali metal atoms may even induce an insulator-to-metal transition,[28] and strong alkali doping may induce structural transformations between different polytypes,[29–34] negative electronic compressibility[22] and bandgap renormalization.[12,35,36] Though Li adsorbates tend to react with the TMD,[37,38] more typically alkali dopants adsorb or intercalate, as shown for Na,[23,29,39,40] K,[11,34,35,41–44] Rb[12,22,45] and Cs.[46,47]

Angle-resolved photoemission spectroscopy (ARPES) has been employed to detect 2DEGs in two-dimensional layered materials.[11,22] ARPES is a powerful tool that measures momentum-resolved electronic structure with band dispersion profiles that may contain identifiable quantum well structures, a key indicator of 2DEGs. The reduced dimensionality of 2DEGs can support high carrier mobility, which plays a pivotal role in high-speed electronic devices, such as field-effect transistors[48] and optoelectronic devices.[49,50] In order to gain insight into the many-body physics at play in these low-dimensional systems, ultrafast carrier dynamics have been particularly powerful.[51–55] In the case of TMDs however, the vast majority of reports on ultrafast carrier dynamics determine quasiparticle lifetimes using time-resolved optical[56–60] and photoemission[61–63] spectroscopy methods that are limited by the intrinsic time resolution available from the laser sources, which is typically tens of femtoseconds. New insights may be gained with core-hole clock spectroscopy (CHC) which overcomes this time restriction by operating in a competitive kinetics scheme, providing a time base intrinsic to the decay lifetime of a core-electron.[64–68] By the very nature of the CHC technique, element and orbital information are selectively encoded in this measurement, which may offer atomistic insights into the carrier dynamics in the 2DEG-like states.

Here, we demonstrate by a combination of ARPES and CHC the effect of K-doping induced massive electronic reconstruction and find dramatically increased electron scattering rates at $\bar{K}$ in the metallic phase of $MoS_2$. The atom-specificity and orbital-selectivity of CHC spectroscopy reveals that the origin for these changes rests in intralayer orbital rehybridization, increasing quantum confinement of the conduction band wavefunction and creating a quasi-freestanding monolayer-like structure of $MoS_2$. We demonstrate a dramatic reduction of the lifetime by a factor of 3 upon K intercalation, and estimate that this corresponds to a lifetime of a few hundred attoseconds. Our study therefore directly demonstrates how reduced screening in the 2DEG-like state of quasi-2D MoS$_2$ increases carrier scattering rates on attosecond time-scales.

*2. Results*

**2.1. K-induced electronic structure effects of bulk MoS$_2$**

The occupied band structure of bulk MoS$_2$ has been well established in both high-resolution angle-resolved photoemission spectroscopy (ARPES)[24] and in state-of-the-art computations.[69] Our data, shown in Fig. 1a along the $\bar{\Gamma} - \bar{K}$ direction, agree well with previous work.[24] We confirm that the valence band maximum (VBM) is located at $\bar{\Gamma}$, and based on the computational studies indicate that the bandgap is indirect and located at $\bar{\Gamma} - \bar{\Sigma}$ with an energy of $E_{g,indirect} = 1.4$ eV.[26] The direct bandgap is at $\bar{K}$, with a magnitude of 1.9 eV.[24,26] Due to the dominant Mo character at the band edge, the valence band supports a spin-orbit splitting of 180 meV at $\bar{K}$, which is not clearly resolved in our room temperature spectra and at the available instrumental resolution; instead we observe a shoulder feature.

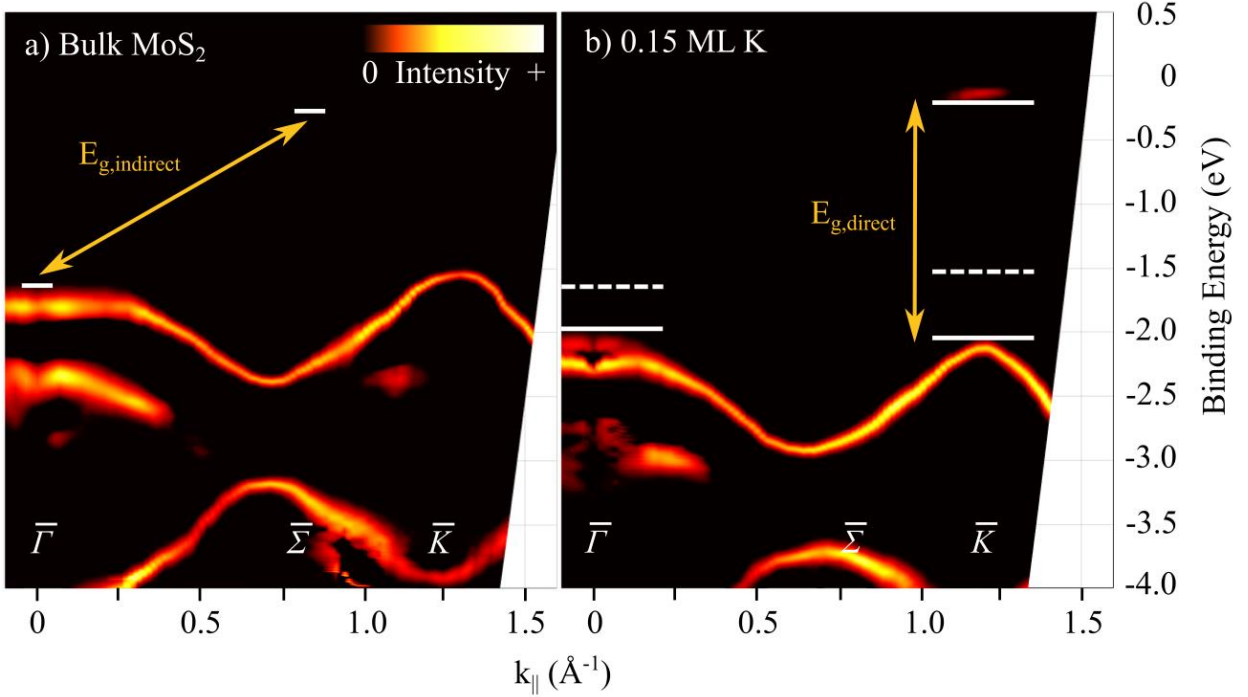

Figure 1. Band structure of as-cleaved and K-dosed bulk MoS2. Band dispersion maps along the $\bar{\Gamma} - \bar{K}$ direction processed with the ARPES curvature method[70] of a) bulk MoS2 and b) 0.15 ML K on bulk MoS2. Filling of the conduction band edge at $\bar{K}$ reveals the direct bandgap, $E_{g,direct} = 2.0(1)$ eV in K-dosed MoS2; whereas the indirect bandgap at $\bar{\Gamma} - \bar{\Sigma}$ ($E_{g,indirect}$) was not observed in either instance. K addition also resulted in differential band shifts at high symmetry points $\bar{\Gamma}$ and $\bar{K}$ represented by solid and dashed lines in the band dispersion maps.

The addition of 0.15 ML K alters the valence band structure, most conspicuously from a semiconducting phase to the known metallic phase with the filling of the conduction band minimum at $\bar{K}$ as illustrated in Fig. 1b, where an electron pocket emerges just below the Fermi level. Additionally, all energy levels shift to higher binding energies represented by dashed to solid lines at the band edges. Remarkably, the bands do not shift rigidly, as might be expected from K-induced *n*-doping. Rather we observe differential shifts: At $\bar{\Gamma}$, the binding energy increases by 0.31(2) eV, while at $\bar{K}$ the shift is 0.72(2) eV. This is a first indication that K atoms do not merely function as *n*-dopants but rather induce a reconstruction of the electronic structure in MoS2. We further observe occupation of the conduction band minimum at $\bar{K}$, creating an approximately direct bandgap system with a bandgap $E_{g,direct} = 2.0(1)$ eV, as measured from the onsets of both bands.

This is another sign of severe electronic reconstruction upon K-doping, since in bulk MoS$_2$ the conduction band minimum is located at $\bar{\Sigma}$ instead of $\bar{K}$. Though the specific changes to the band structure differ somewhat on a quantitative level, e.g. in the magnitude of the shifts of band extrema at $\bar{\Gamma}$ and $\bar{K}$, the transition to a direct-gap system and the accompanying electronic reconstruction are reminiscent of electronic structure differences between bulk and monolayer MoS$_2$.[24,26] In that case, band structure changes are generally believed to be due to quantum-confinement of out-of-plane orbitals at $\bar{\Gamma}$ and $\bar{\Sigma}$ while leaving the Mo $d_{x^2-y^2}$ and $d_{xy}$ in-plane orbitals that dominate the valence band at $\bar{K}$ unaffected.[71] Addition of potassium atoms to *bulk* MoS$_2$ therefore induces a measure of layer decoupling and the emergence of MoS$_2$ structures that are rather similar to nominally isolated monolayers. True MoS$_2$ monolayers are normally only accessible by exfoliation or direct growth. Importantly, this decoupling is accompanied by filling of the $\bar{K}$-valley by electrons, creating a state that resembles and approximates a multivalley 2DEG, as also suggested for Rb-dosed WSe$_2$.[22]

The majority of K atoms initially adsorb on the surface, followed by prompt intercalation into the layered MoS$_2$ crystal, as observed in the XPS data (Supp. Note 1 and Supp. Fig. 1). This is supported by both work function and band structure measurements that lack large time-dependent changes, remaining stable over many hours after dosing, and the fact that the K coverages and a temperature of 298 K used here kinetically favor intercalation.[46,72,73] Indeed, surface adsorption has been reported to shift the band structure rigidly,[74] in contrast to our findings. K intercalation is also supported by theoretical predictions[75,76] and previous electronic structure investigations.[35]

Our ARPES results are in good agreement with other reports on alkali dosing of TMDs. The strong electronic reconstruction upon K doping leads to orbital rehybridization, with the K-doped MoS$_2$ resembling monolayer MoS$_2$ with enhanced 2D confinement and resembling the formation of a 2DEG. ARPES does however not provide a microscopic mechanism for this effect, and we turn instead to element-selective resonant X-ray photoemission spectroscopy.

**2.2. Carrier dynamics of 2DEG-like state in K-doped bulk MoS$_2$**

As a first step towards tracking the conduction band dynamics, we investigate the X-ray absorption (XA) spectra of as-cleaved and K-dosed MoS$_2$ on the Mo M$_{2,3}$ adsorption edges (primarily $3p \rightarrow 4d$) to map the desired conduction band region (Fig. 2). Upon K intercalation, the intensity decreases and shifts by 0.39(2) eV towards lower photon energies, but is not accompanied by broadening or peak shape changes (Supp. Note 2 and Supp. Fig. 2). The change of the XA spectrum is likely due to filling of the conduction band states, thus lowering the unoccupied density of states and reducing the screening of the core-hole.

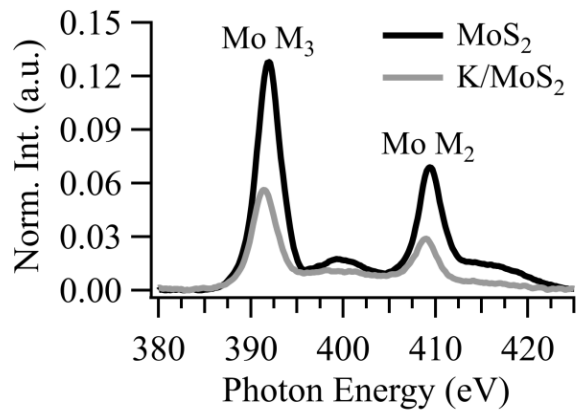

Figure 2. X-ray absorption spectra of cleaved (black) and K-dosed (gray) MoS$_2$ monitoring Mo M$_{2,3}$ absorption edges. Intensity is normalized with respect to the incident X-ray photon flux, $I_0$. Each XA spectrum shows the spin-orbit components of Mo M$_3$ ($3p_{3/2}$) and M$_2$ ($3p_{1/2}$).

The conduction band of MoS$_2$ is composed of strongly hybridized Mo $4d \otimes$ S $3p \otimes$ S $3s$ orbitals, with contributions from Mo $5p$ as well, as determined by polarization-dependent XAS investigations on the Mo L$_{2,3}$ edge by Li *et al.*,[77] and from group theoretical considerations.[78] At $\bar{\Gamma}$ the conduction band is comprised mainly of Mo $d_{x^2-y^2}$ and $d_{xy}$ orbitals, hybridized with S $p_x$, $p_y$ orbitals, while at $\bar{K}$ and $\bar{M}$ the lowest few conduction bands support out-of-plane $d$ and $p$ character oriented along the $z$-axis (surface normal).

To shed more light on the nature of the 2DEG-like state and the massive electronic reconstruction resulting from the formation of the metallic phase upon K intercalation, we monitor the decay of the excited state electron using core-hole-clock spectroscopy (CHC). CHC is a rather general approach in which carrier lifetimes are retrieved from resonant photoemission spectroscopy *intensity* measurements.[64] Specifically, while scanning over the Mo $M_{2,3}$ absorption edge and following resonant excitation of the core electron to the conduction band (Mo $3p \rightarrow 4s, 4d$), we monitor the photoemission intensity associated with specific decay channels of the core-excited electron from the conduction band. Comparison of this intensity in $MoS_2$ vs. $K/MoS_2$, in relation to the known lifetime of the Mo $3p$ core hole, enables us to obtain excited state dynamics with element- and orbital-specificity and on attosecond time-scales, as discussed in what follows.[79]

The scheme used here is shown in Fig. 3: Resonant excitation from Mo $3p$ levels to the conduction band of $MoS_2$ (Fig. 3a) is followed by an Auger-like decay that results in a one-hole final state in the Mo $3d$ manifold (Fig. 3b). In the CHC experiment, this participator decay channel manifests as resonant enhancement of photoemission from Mo $3d$ levels. If on the time-scale of the Auger decay lifetime the excited electron escapes the potential exerted by the core hole e.g. due to delocalization or electron-electron scattering, the expected resonant photoemission enhancement is however suppressed (Fig. 3c). The excited state decays instead via normal Auger processes.[64] Consequently, Mo $3d$ photoemission appears with an intensity comparable to direct photoemission at similar photon energies. Resonant photoemission intensities therefore directly encode the dynamics of carriers.

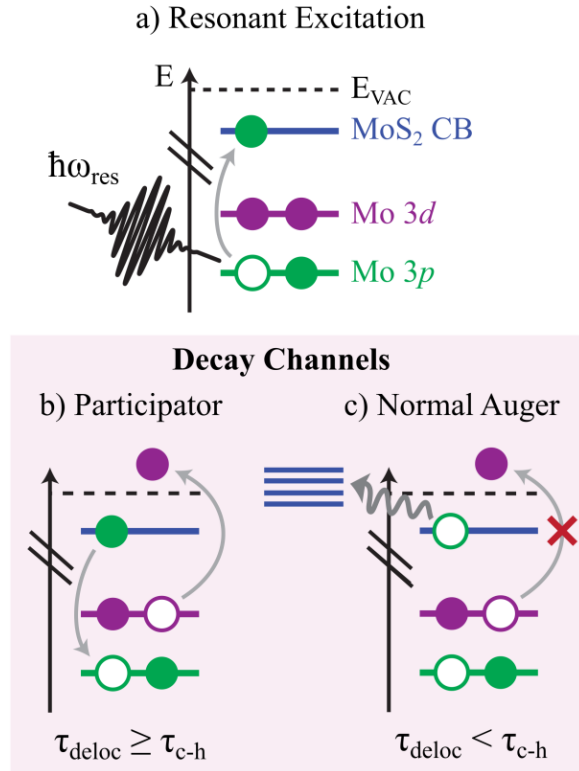

Figure 3. CHC scheme for measuring attosecond conduction band dynamics in MoS$_2$ following the decay channels after resonant excitation. a) Resonant excitation from Mo **$3p$** (green) to the conduction band of MoS$_2$ (blue) results in a highly excited state that decays via two types of Auger-like pathways: b) participator decay resulting in a final state with Mo **$3d$** hole (purple) or c) the excited electron delocalizes into the conduction band before core-hole decay, and normal Auger processes drive the system to its ground state. Note that $\boldsymbol{\tau_{c-h}}$ is the core-hole lifetime and $\boldsymbol{\tau_{CB}}$ is the lifetime of the conduction band electron.

We first consider the resonant photoemission (RPE) spectra whereby we can ultimately extract the carrier dynamics from the intensity fluctuations in the observed resonantly excited XPS features. Figure 4 shows RPES maps of freshly cleaved MoS$_2$ (Fig. 4a) and K-dosed MoS$_2$ (Fig. 4b), together with the XA spectrum of Mo M$_{2,3}$ (Fig. 4c) for convenience. The RPES map is comprised of X-ray photoemission (XP) spectra of Mo $3d_{5/2}$, Mo $3d_{3/2}$ and S $2s$ features at photon energies taken in the vicinity of the Mo M$_{2,3}$ absorption edge and measured every 0.5 eV. Accessibility to the MoS$_2$ conduction band results in strong resonant enhancement of both Mo $3d_{3/2}$ and Mo $3d_{5/2}$ XP features. Note that the photoemission intensity from the S 2s core-level is too weak to reliably determine the occurrence of resonance enhancement. In the remainder of the results section and for a quantitative analysis, we focus on the intensity changes over the Mo $M_3$ adsorption edge to avoid interference from the K N-edge at 410 eV that overlaps with the Mo $M_2$ adsorption edge.

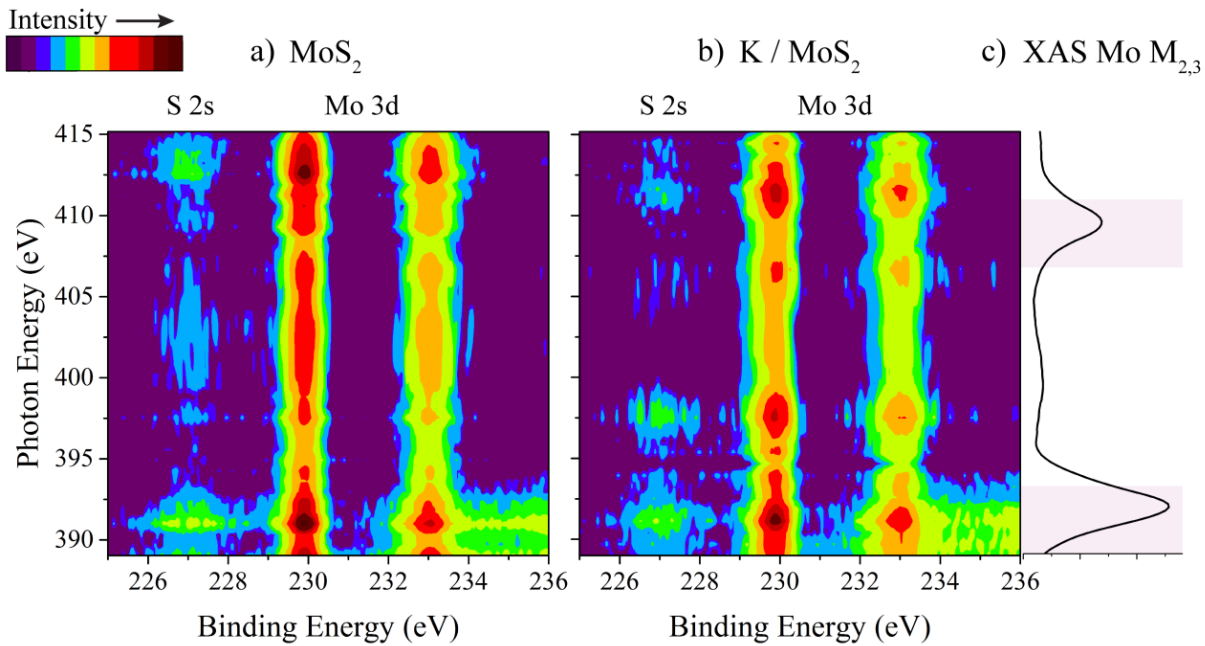

Figure 4. Resonant photoemission contour plots of a) as-cleaved and b) K-dosed $MoS_2$ monitoring the intensity changes of XP features Mo $\mathbf{3d}$ and S $\mathbf{2s}$ on resonance with Mo $M_{2,3}$. c) XA spectrum of Mo $M_{2,3}$ with highlighted regions (pink) where resonantly enhanced photoemission may be expected in the RPES maps.

Figure 5 features representative XP spectral cuts from the RPES maps for both $MoS_2$ (Fig. 5a) and K-dosed $MoS_2$ (Fig. 5b) at a cut along $\hbar\omega = 391$ eV. All XPS fits incorporate features for relevant core-levels including Mo $3d_{5/2}$, Mo $3d_{3/2}$ and S $2s$ as well as an Auger feature, identified by the fact that it shifts binding energy with a change in photon energy. Notably, the spin-orbit split Mo $3d$ peak shapes change from symmetric to asymmetric with the introduction of K. The peak shape is reminiscent of the classical Doniach-Sunjic asymmetry resulting from metallic screening, which might be expected to arise in the transition from the semiconducting to the metallic phase of $MoS_2$ upon K dosing. However, the many-body processes giving rise to a

Doniach-Sunjic line profile result in an asymmetry extending towards higher binding energies, opposite to what is observed in Fig. 5. We instead assign a shoulder feature to the fit that represents a small fraction of partially reduced $MoS_2$, indeed also expected from the metallization of $MoS_2$.

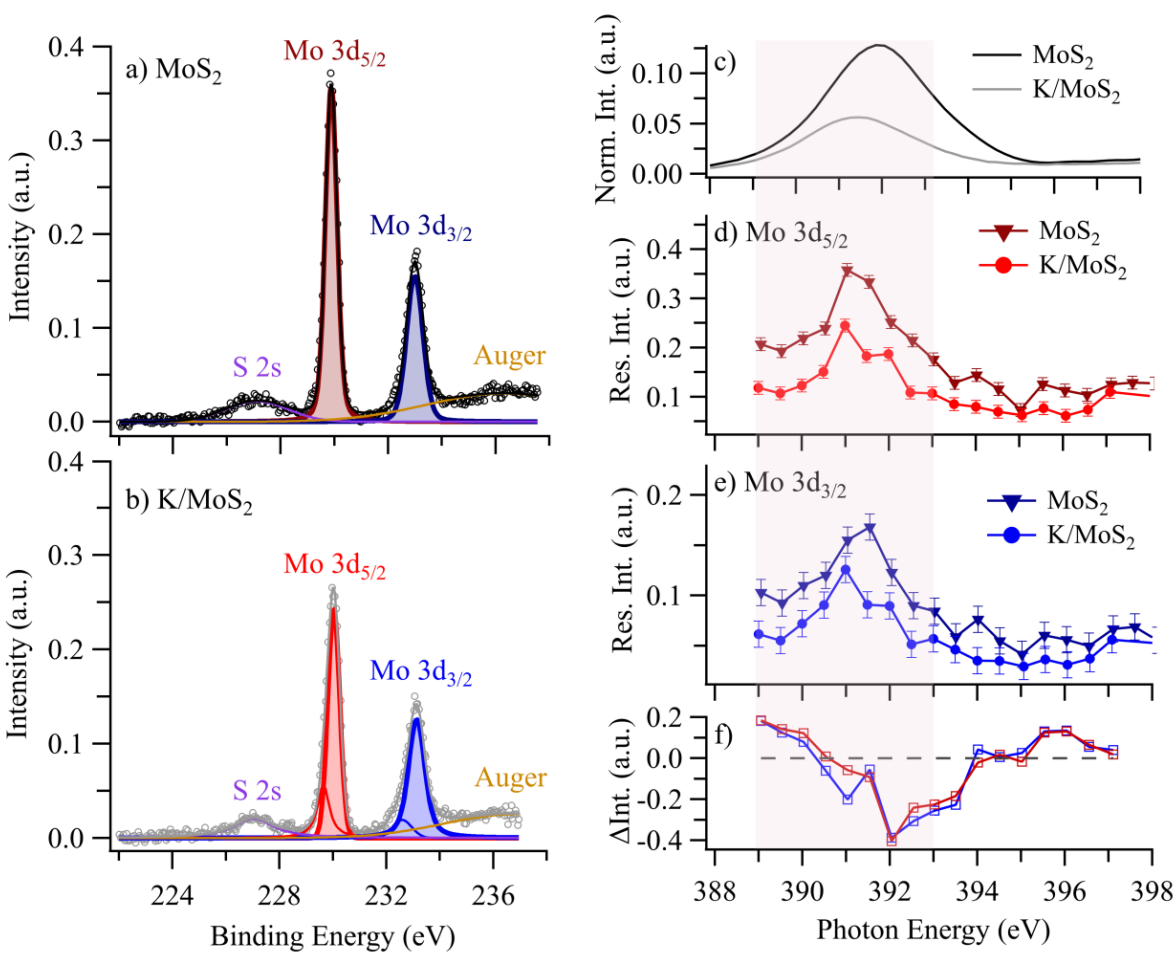

Figure 5. Resonant enhancement of Mo $3d$ core-levels upon excitation into the $MoS_2$ conduction band for as-cleaved and K-dosed $MoS_2$. Representative spectra at $\hbar\omega = 391$ eV for a) $MoS_2$ and b) K/$MoS_2$. Apart from intensity changes, a small contribution of reduced Mo accounts for the slightly asymmetric peak shape in K/$MoS_2$, not observed in $MoS_2$. c) The XA spectra of the Mo $M_3$ adsorption edge. Normalized resonant intensities derived from XP spectral fits of $MoS_2$ and K/$MoS_2$ for d) Mo $3d_{5/2}$ and e) Mo $3d_{3/2}$ core-levels. The error bar represents uncertainties in the X-ray intensity, obtained from Au $4f_{7/2}$ photoemission spectra acquired concurrently. f) Difference of normalized resonant and XA intensities as a function of photon energy that reveals additional changes in resonant intensities beyond conduction band filling.

From the XPS fits, we extract the normalized photoemission intensities of the Mo $3d$ features as a function of photon energy across the Mo $M_3$ edge (Fig. 5c), excluding the minor contribution from Mo reduction in K/MoS$_2$ XP spectra. The resulting resonant intensities are shown for Mo $3d_{5/2}$ in Fig. 5d and for Mo $3d_{3/2}$ in Fig. 5e. Two salient features are noticeable: 1) The photoemission intensities exhibit a clear resonant enhancement on the Mo $M_3$ edge which does however not follow the complete XAS profile; and 2) K-dosing reduces the resonant intensity for both Mo $3d$ photoemission channels.

The question arises if this behavior is merely a consequence of conduction band filling. To this end, we calculate as a function of photon energy the difference $\Delta Int$ between normalized resonant intensities in MoS$_2$ ($I_{MoS_2}^{Res}$) and K/MoS$_2$ ($I_{K,MoS_2}^{Res}$) and normalized XA intensities in MoS$_2$ ($I_{MoS_2}^{XAS}$) and K/MoS$_2$ ($I_{MoS_2}^{XAS}$) and plot it in Fig. 5f:

$$\Delta Int = \left[\frac{I_{MoS_2}^{Res} - I_{K,MoS_2}^{Res}}{I_{MoS_2,max}^{Res}}\right] - \left[\frac{I_{MoS_2}^{XAS} - I_{K,MoS_2}^{XAS}}{I_{MoS_2,max}^{XAS}}\right] \quad (1)$$

If conduction band filling was the dominant reason for the suppression of resonant photoemission intensity upon dosing with K, the values for $\Delta Int$ would be scattered around 0 (dashed line in Fig. 5f). The systematic negative deviation from $\Delta Int = 0$ indicates clearly that conduction band filling is not primarily responsible for the much-reduced resonant photoemission intensity in K/MoS$_2$. Rather, the reduced resonant intensities in both photoemission channels report on a shortened excited state lifetime upon K intercalation and formation of a 2DEG-like state. Averaged across the absorption edge, this intensity reduction amounts to a factor of $1.69 \pm 0.09$ for Mo $3d_{5/2}$ and $1.55 \pm 0.17$ for Mo $3d_{3/2}$, and constitutes our principal experimental finding. We note that the reduced intensity is not merely a reflection of the limited escape depth of the Auger

14electrons, since both with and without K the resonant photoemission experiments probe primarily the top layer of the crystal at the relevant kinetic energies.

We may estimate the drop in lifetime of a carrier in the conduction band by making use of the known Mo $3p_{3/2}$ core-hole lifetime of $\tau_{c-h} = \frac{1}{k_{c-h}} = 299$ as.[80] Since the hole is an inner-shell core-hole and thus well screened, its lifetime is independent of the chemical environment, a fact we will use below to estimate absolute lifetimes.[64] The core-hole functions as a natural clock against which electron dynamics in the conduction band can be measured in a competitive kinetics scheme (Fig. 3, Supp. Note 3 and Supp. Fig. 3): The competition of core-hole decay vs. delocalization and scattering processes in the conduction band determines the intensity of the observed Mo $3d$ participator. We conservatively estimate that a signal-to-noise ratio of 10 is needed to differentiate intensities, which translates to an observable range of carrier lifetimes of $0.1\tau_{c-h} \leq \tau_{CB} \leq 10\tau_{c-h}$, where $\tau_{CB} = \frac{1}{k_{CB}}$ is the lifetime of carriers in the conduction band. We thus expect to be able to observe processes on time-scales from 30 attoseconds to 3.0 femtoseconds. Assuming first-order rate laws for all decay processes, as is customary for CHC, $\tau_{CB}$ is obtained from the measured resonant photoemission intensities of the Mo $3d$ participator in MoS$_2$ ($I_{MoS_2}$) and the resonant photoemission intensity associated with a fully localized conduction band electron ($I_{loc}$) that undergoes exclusively core-hole decay:[64]

$$k_{CB} = k_{c-h} \frac{I_{loc} - I_{MoS_2}}{I_{MoS_2}} \qquad (2)$$

Conduction band electrons in neither MoS$_2$ nor K/MoS$_2$ may be assumed to be fully localized, and hence $I_{loc}$ is not directly accessible from our measurement. However, we may anchor a lifetime measurement with CHC from estimating $I_{loc}$ by making use of the known conduction band carrier lifetime of bulk MoS$_2$ of 1.25 fs, obtained by CHC in multi-layer MoS$_2$.[81]

This allows us to estimate $I_{loc}$ of a putative fully localized Mo $3d_{3/2}$ participator from Eqn. (2). We will discuss the validity of this assumption below, but note at present that the lifetime is also consistent with the width of the XA spectrum. The resonant photoemission intensities for the Mo $3d_{5/2}$ and $3d_{3/2}$ participator in Fig. 5d,e are then used to extract the carrier lifetimes for MoS$_2$ and K/MoS$_2$ excited on the Mo M$_3$ edge, as shown in Fig. 6 and with peak values tabulated in Table 1.

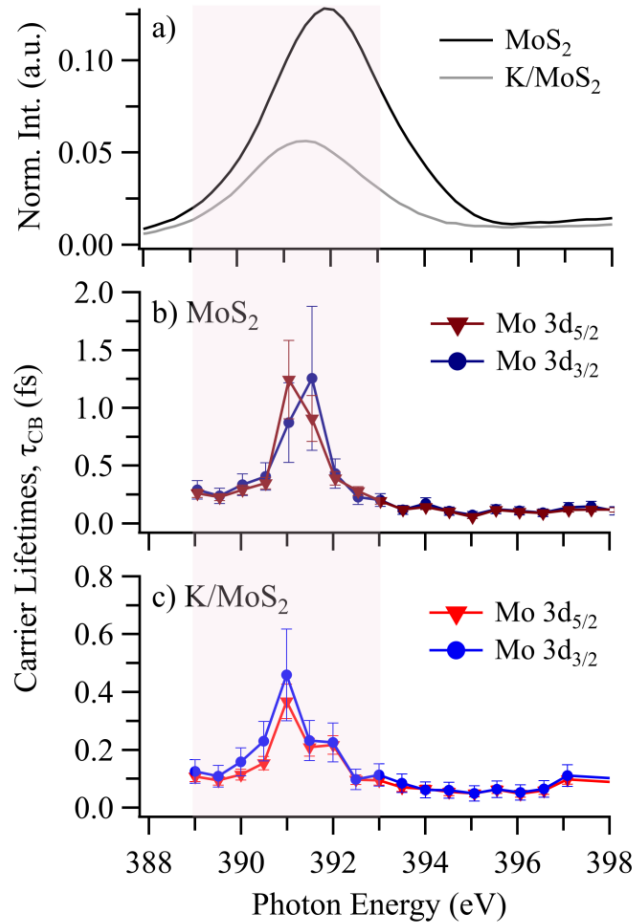

Figure 6. Carrier lifetimes of as-cleaved and K-dosed MoS$_2$. a) Mo M$_3$ XA spectrum. Carrier lifetimes of b) MoS$_2$ and c) K/MoS$_2$ calculated from Mo $3d$ resonant intensities and their uncertainties, assuming $\tau_{CB} = 1.24$ fs for MoS$_2$.

Table 1. Ultrafast carrier delocalization times ($\tau_{CB} = k_{CB}^{-1}$) of MoS$_2$ and K/MoS$_2$ at peak values from Fig. 6 at approximately $\hbar\omega$ = 391 eV.

| XP feature | $\tau_{CB}$, MoS$_2$ (fs) | $\tau_{CB}$, K/MoS$_2$ (as) |
| --- | --- | --- |
| Mo $3d_{5/2}$ | $1.24 \pm 0.34$ | $368 \pm 60$ |
| Mo $3d_{3/2}$ | $1.25 \pm 0.62$ | $459 \pm 159$ |

The effect of K intercalation and formation of a 2DEG-like state is striking: The carrier lifetimes drop from approximately 1.25 fs to 300 – 500 as in K-dosed MoS$_2$, approximately 3 × shorter than for MoS$_2$, and indicating that K intercalation generates additional pathways for inelastic scattering of conduction band electrons. Both participator channels, Mo $3d_{5/2}$ and Mo $3d_{3/2}$, exhibit a similar decrease, where the difference stems largely from different energy-dependent partial photoionization cross-sections.[82]

*3. Discussion*

We used the known $\tau_{CB}$ of MoS$_2$ to anchor our lifetime estimates.[81] This value was obtained for a measurement on the S K-edge rather than the Mo M-edge, and some caution is in order whether this value would be the same when measured on the Mo M-edge. In a recent resonant Auger study of c(4x2)S / Ru(0001) probing core-holes on different edges ($\Delta(\hbar\omega)$ ~2250 eV), comparable carrier delocalization times were found, independent of the X-ray absorption edge.[83] This is reasonable if the core-hole is located significantly below the valence shell and therefore well shielded by the remaining electrons. This is certainly also the case for the Mo $3p$ levels used here, and one may thus expect that 1.25 fs is a reasonable estimate for the lifetime in the conduction band. If instead we vary this lifetime between 100 as and 100 fs for a given change in resonant photoemission intensity of approximately 1.5, we find that the lifetime in K/MoS$_2$ drops between a factor 1.6 and 160. Consequently, K intercalation and the accompanying transition to the metallic phase of MoS$_2$ leads in all cases to drastically reduced lifetimes. We note in passing that these lifetimes are measured in the presence of a core-hole, and hence should not be compared with quasiparticle lifetimes that might be estimated from ARPES bandwidths.

The combination of ARPES and CHC spectroscopy reveals that K intercalation causes a strong electronic reconstruction, accompanied by the formation of a 2DEG-like state and significantly shorter electronic lifetimes in the conduction band. The element- and orbital-selectivity of XPS help distinguish between different possible mechanisms at the root of these strong electronic changes.

A number of studies have demonstrated that alkali atom dosing induces a structural phase transition from the standard *2H*- to *1T*-MoS$_2$, triggered by intralayer atomic gliding.[33,84] Signatures of this transition to the metallic *1T* phase can be readily observed in XPS, manifesting e.g. as shifts in the Mo $3d$ core-level by several eV.[85] We do not observe such changes in XPS of Mo $3d$. Moreover, though charge-density wave formation and suppression by alkali adsorbates has been observed in other van der Waals layered materials, there is no indication of such a phase transition in the ARPES data.[45] We may therefore exclude structural phase transformations as the source of the 2D-like confinement.

Previous work showed that layered van der Waals materials are effectively two-dimensional on short time-scales, i.e. intralayer electron-transfer is approximately $10 \times$ faster than interlayer electron transfer, $k_{intra} \approx 10 k_{inter}$.[66] It is conceivable that K intercalation further separates adjacent MoS$_2$ sheets, thus further decoupling individual layers. Such simple "layer decoupling" is however unable to account for the observed shortened lifetimes upon K intercalation: Under the assumption of first-order kinetics inherent to CHC, and assuming the two rates $k_{intra}$ and $k_{inter}$ to be independent, the total rate $k_{tot}$ for conduction band electron transfer in real space is $k_{tot} = k_{intra} + k_{inter}$; simple layer decoupling due to increased separation of MoS$_2$ sheets by K intercalation would reduce $k_{inter}$ and therefore decrease $k_{tot}$, contrary to our observations of shortened carrier lifetimes and increased $k_{CB}$.

Instead, we make use of the element- and orbital-specificity of CHC spectroscopy to identify the mechanism for formation of the 2DEG-like state. The *p*-polarized X-ray radiation probes predominantly out-of-plane Mo orbital constituents of the conduction band. In combination with the atomic selection rule of $\Delta \ell = \pm 1$, excitation from the Mo $3p$ core levels accesses preferentially out-of-plane $d$-orbitals such as $4d_{z^2}$. A group theory-based orbital decomposition of the MoS$_2$ conduction band indicates that such orbitals occur primarily at high symmetry point $\bar{K}$ and $\bar{M}$, and thus CHC spectroscopy preferentially probes electron dynamics in these two regions.[78] The $\bar{K}$ valley or conduction band minimum is also the region in the Brillouin zone of increased electron density upon K intercalation, and hence the most significant region for understanding the electronic reconstruction and the associated conduction band electron dynamics.

The out-of-plane character of the orbitals probed in our experiment accesses Mo orbitals that are hybridized with S ligand orbitals. Thus, an increase of $k_{CB}$ is the result of rehybridization and stronger mixing between Mo $4d_{z^2}$ and S $3p$ orbitals in K-dosed MoS$_2$. In real space, this results in increased electron-electron interactions, i.e. increased scattering rates, *within a single MoS$_2$ layer* (increased $k_{intra}$), and provides additional decay channels of the excited state. The decreased electronic lifetime in electron-doped MoS$_2$ involves both an increased phase-space for electron scattering and an increased scattering cross-section. The electron-doping induced many-body interactions cause rehybridization towards an electronic structure that resembles more closely that of the MoS$_2$ monolayer, which is the source of the strong electronic reconstruction and appearance of a direct bandgap in the ARPES map. At the same time, the increased intralayer interactions occur at the expense of interlayer coupling, confining the electrons more strongly to a single MoS$_2$ sheet and creating the observed 2DEG-like state.

*4. Conclusions*

We demonstrate with a combined ARPES and CHC study that K intercalation in $MoS_2$ leads to massive electronic reconstruction and formation of a 2DEG-like state due to out-of-plane orbital rehybridization. The increased conduction band electron density and quantum confinement to a single $MoS_2$ layer causes dramatically increased electronic scattering rates and reduced lifetimes, which we estimate to be reduced by a factor ∼3 to approximately 300 – 500 attoseconds. Such fast electron scattering is a hallmark of reduced screening that accompanies confinement of layered van der Waals materials in the 2D limit. Evidently, this confinement-induced reduced screening is not offset by the doping-induced increased carrier density, and is rather the result of many-body interactions induced by filling of the $\bar{K}$ valley upon electron-doping. Our study also shows that the ultrafast carrier dynamics of $MoS_2$ can be tailored rather simply, and opens opportunities for investigating the consequences of intercalation- or adsorption-induced inversion symmetry breaking and the possibility of spin-splitting on ultrashort time-scales.[12]

*Methods*

Synthetically grown bulk 2H-$MoS_2$ was purchased from 2D Semiconductors. The $MoS_2$ crystal was cleaved by mechanical exfoliation in a protective atmosphere before introduction to an ultrahigh vacuum preparation chamber. Sample integrity was verified by XPS of Mo $3d_{5/2}$ and $3d_{3/2}$ core-levels, sharp and intense valence band features without an underlying defect background, and a work function of 5.2 eV. K deposition (SAES Getters) was calibrated on clean Cu(111), where the evolution of the electronic structure as a result of alkali adsorbates has been studied in detail, and is reported in fractions of a putative monolayer (ML).[86–88]

XA and RPE spectra were obtained at Stanford Synchrotron Radiation Lightsource on beamline 8-2 (SLAC National Accelerator Laboratory), with all experiments carried out at room temperature. XA spectra were performed in total electron yield (TEY) mode using a drain current. XP and RPE spectra were acquired with a double-pass cylindrical mirror analyzer for kinetic energy discrimination of photoelectrons. The electric field of the near-grazing incidence X-ray beam was linearly polarized in the horizontal plane, and excites predominantly transitions that are polarized along surface normal (*z*-axis). XP and RPE spectral resolution was approximately 0.5 eV, with a pass energy of 25 eV. The base pressure of the analysis chamber remained below 7 x $10^{-9}$ mbar throughout all measurements. Absolute energy calibration and global vacuum level corrections were obtained from an Au film on a Si wafer, and spectral intensities were calibrated against the incident X-ray photon flux ($I_0$) measured on a gold grid upstream from the analysis chamber. All XP spectra were processed with an initial subtraction of a linear background and a Shirley or integrated background.

ARPE spectra were collected in a VG EscaLab MK II photoelectron spectrometer at room temperature with a base pressure of 2 x $10^{-10}$ mbar. All reported spectra were measured along the $\bar{\Gamma} - \bar{K}$ direction with an acceptance angle of ±1.5° and a sample bias of -3 V using a He (I) source (SPECS 10/35, $\hbar\omega$ = 21.22 eV) mounted at a 30° angle of incidence. ARPES data were processed using the curvature method.[70]

*Acknowledgments*

This work was supported by the National Science Foundation, award numbers ECCS 1708652. Use of the Stanford Synchrotron Radiation Lightsource, SLAC National Accelerator Laboratory, is supported by the U.S. Department of Energy, Office of Science, Office of Basic

Energy Sciences under Contract No. DE-AC02-76SF00515. The authors thank A. Batyrkhanov for help with data processing.

*Contributions*

C.N.E. and S.L.Z. carried out the experiments, C.N.E., J.P. and A.C. analyzed the results, and O.L.A.M. conceived and designed the experiments. C.N.E., D.L.N. and O.L.A.M. wrote the manuscript and discussed the results.

*Competing interests*

The authors declare no competing financial interests.

*References*

1. Ohtomo, A. & Hwang, H. Y. A high-mobility electron gas at the LaAlO3/SrTiO3 heterointerface. *Nature* **427**, 423–426 (2004).

2. Thiel, S., Hammerl, G., Schmehl, A., Schneider, C. W. & Mannhart, J. Tunable quasi-two-dimensional electron gases in oxide heterostructures. *Science (80-. ).* **313**, 1942–1945 (2006).

3. Reyren, N. *et al.* Superconducting interfaces between insulating oxides. *Science* **317**, 1196–9 (2007).

4. Brinkman, A. *et al.* Magnetic effects at the interface between non-magnetic oxides. *Nat. Mater.* **6**, 493–496 (2007).

5. Störmer, H. L., Dingle, R., Gossard, A. C., Wiegmann, W. & Sturge, M. D. Two-dimensional electron gas at a semiconductor-semiconductor interface. *Solid State Commun.* **29**, 705–709 (1979).

6. Abstreiter, G., Brugger, H., Wolf, T., Jorke, H. & Herzog, H. J. Strain-induced two-dimensional electron gas in selectively doped Si/SixGe1-x superlattices. *Phys. Rev. Lett.* **54**, 2441–2444 (1985).

7. Tsui, D. C., Stormer, H. L. & Gossard, A. C. Two-dimensional magnetotransport in the extreme quantum limit. *Phys. Rev. Lett.* **48**, 1559–1562 (1982).

8. Berger, C. *et al.* Ultrathin epitaxial graphite: 2D electron gas properties and a route toward graphene-based nanoelectronics. *J. Phys. Chem. B* **108**, 19912–19916 (2004).

9. Kanetani, K. *et al.* Ca intercalated bilayer graphene as a thinnest limit of superconducting C6Ca. *Proc. Natl. Acad. Sci. U. S. A.* **109**, 19610–19613 (2012).

10. Li, L. *et al.* Quantum oscillations in a two-dimensional electron gas in black phosphorus

thin films. *Nat. Nanotechnol.* **10**, 608–613 (2015).

11. Alidoust, N. *et al.* Observation of monolayer valence band spin-orbit effect and induced quantum well states in MoX2. *Nat. Commun.* **5**, 4673 (2014).

12. Kang, M. *et al.* Universal Mechanism of Band-Gap Engineering in Transition-Metal Dichalcogenides. *Nano Lett.* **17**, 1610–1615 (2017).

13. Gehlmann, M. *et al.* Quasi 2D electronic states with high spin-polarization in centrosymmetric MoS2 bulk crystals. *Sci. Rep.* **6**, 26197 (2016).

14. Mak, K. F., He, K., Shan, J. & Heinz, T. F. Control of valley polarization in monolayer MoS2 by optical helicity. *Nat. Nanotechnol.* **7**, 494–498 (2012).

15. Bertoni, R. *et al.* Generation and Evolution of Spin-, Valley-, and Layer-Polarized Excited Carriers in Inversion-Symmetric WSe 2. *Phys. Rev. Lett.* **117**, 277201 (2016).

16. Deng, K. *et al.* Experimental observation of topological Fermi arcs in type-II Weyl semimetal MoTe2. *Nat. Phys.* **12**, 1105–1110 (2016).

17. Choe, D.-H., Sung, H.-J. & Chang, K. J. Understanding topological phase transition in monolayer transition metal dichalcogenides. *Phys. Rev. B* **93**, 125109 (2016).

18. Mak, K. F., McGill, K. L., Park, J. & McEuen, P. L. The valley Hall effect in $MoS_2$ transistors. *Science* **344**, 1489–92 (2014).

19. Okamoto, N. *et al.* Electric control of the spin Hall effect by intervalley transitions. *Nat. Mater.* **13**, 932–937 (2014).

20. Schaibley, J. R. *et al.* Valleytronics in 2D materials. *Nat. Rev. Mater.* **1**, 16055 (2016).

21. Wu, Y., Tong, Q., Liu, G.-B., Yu, H. & Yao, W. Spin-valley qubit in nanostructures of monolayer semiconductors: Optical control and hyperfine interaction. *Phys. Rev. B* **93**, 045313 (2016).

22. Riley, J. M. *et al.* Negative electronic compressibility and tunable spin splitting in WSe2. *Nat. Nanotechnol.* **10**, 1043–1047 (2015).

23. Eknapakul, T. *et al.* Nearly-free-electron system of monolayer Na on the surface of single-crystal HfSe2. *Phys. Rev. B* **94**, 201121 (2016).

24. Jin, W. *et al.* Direct Measurement of the Thickness-Dependent Electronic Band Structure of MoS2 Using Angle-Resolved Photoemission Spectroscopy. *Phys. Rev. Lett.* **111**, 106801 (2013).

25. Yeh, P.-C. *et al.* Layer-dependent electronic structure of an atomically heavy two-dimensional dichalcogenide. *Phys. Rev. B* **91**, 041407 (2015).

26. Mak, K. F., Lee, C., Hone, J., Shan, J. & Heinz, T. F. Atomically Thin MoS2: A New Direct-Gap Semiconductor. *Phys. Rev. Lett.* **105**, 136805 (2010).

27. Lezama, I. G. *et al.* Indirect-to-Direct Band Gap Crossover in Few-Layer MoTe2. *Nano Lett.* **15**, 2336–2342 (2015).

28. Ahmad, M. *et al.* Semiconductor-to-metal transition in the bulk of WSe2 upon potassium intercalation. *J. Phys. Condens. Matter* **29**, 165502 (2017).

29. Brauer, H. E., Starnberg, H. I., Holleboom, L. J., Hughes, H. P. & Strocov, V. N. Modifying the electronic structure of TiS2 by alkali metal intercalation. *J. Phys. Condens. Matter* **11**, 8957–8973 (1999).

30. Pronin, I. I., Gomoyunova, M. V., Valdaitsev, D. A. & Faradzhev, N. S. Initial stages in the intercalation of 1T-TiS2(0001) single crystals by potassium. *Phys. Solid State* **43**, 1788–1793 (2001).

31. Brauer, H. E., Starnberg, H. I., Holleboom, L. J. & Hughes, H. P. In situ intercalation of the layered compounds TiS2, ZrSe2 and VSe2. *Surf. Sci.* **331**–**333**, 419–424 (1995).

32. Remškar, M., Popović, A. & Starnberg, H. I. Stacking transformation and defect creation in Cs intercalated TiS2 single crystals. *Surf. Sci.* **430**, 199–205 (1999).

33. Papageorgopoulos, C. A. & Jaegermann, W. Li intercalation across and along the van der Waals surfaces of MoS2(0001). *Surf. Sci.* **338**, 83–93 (1995).

34. Rossi, A. *et al.* Phase transitions via electron doping in WTe2. (2020).

35. Eknapakul, T. *et al.* Electronic Structure of a Quasi-Freestanding MoS2 Monolayer. *Nano Lett.* **14**, 1312–1316 (2014).

36. Kim, B. S. *et al.* Possible electric field induced indirect to direct band gap transition in MoSe2. *Sci. Rep.* **7**, 1–6 (2017).

37. Kamaratos, M. *et al.* Interaction of Li with the group IV selenide layer compounds at low temperature. *J. Phys. Condens. Matter* **14**, 8979–8986 (2002).

38. Schellenberger, A., Jaegermann, W., Pettenkofer, C., Kamaratos, M. & Papageorgopoulos, C. A. Li insertion into 2H-WS2: Electronic structure and reactivity of the UHV In-situ prepared interface. *Berichte der Bunsengesellschaft für Phys. Chemie* **98**, 833–841 (1994).

39. Komesu, T. *et al.* Adsorbate doping of MoS 2 and WSe 2 : the influence of Na and Co. *J. Phys. Condens. Matter* **29**, 28 (2017).

40. Yakovkin, I. N. Band structure of the MoS2 bilayer with adsorbed and intercalated Na. *Phys. status solidi* **252**, 2693–2697 (2015).

41. Pronin, I. I., Gomoyunova, M. V., Faradzhev, N. S., Valdaitsev, D. A. & Starnberg, H. I. In-situ intercalation of VSe2(0001) with K: direct observation of near-surface structure transformation by incoherent medium-energy electron diffraction. *Surf. Sci.* **461**, 137–145 (2000).

42. Zhang, Y. *et al.* Direct observation of the transition from indirect to direct bandgap in atomically thin epitaxial MoSe2. *Nat. Nanotechnol.* **9**, 111–115 (2014).

43. Zhang, Y. *et al.* Electronic Structure, Surface Doping, and Optical Response in Epitaxial WSe2 Thin Films. *Nano Lett.* **16**, 2485–2491 (2016).

44. Miwa, J. A. *et al.* Electronic structure of epitaxial single-layer MoS2. *Phys. Rev. Lett.* **114**, 046802 (2015).

45. Rossnagel, K. Suppression and emergence of charge-density waves at the surfaces of layered 1T-TiSe2 and 1T-TaS2 by in situ Rb deposition. *New J. Phys.* **12**, 125018 (2010).

46. Starnberg, H. I., Brauer, H. E. & Strocov, V. N. Low temperature adsorption of Cs on layered TiS2 studied by photoelectron spectroscopy. *Surf. Sci.* **384**, L785–L790 (1997).

47. Sant, R. *et al.* Decoupling Molybdenum Disulfide from Its Substrate by Cesium Intercalation. *J. Phys. Chem. C* (2020). doi:10.1021/acs.jpcc.0c00970

48. Cheng, R. *et al.* Few-layer molybdenum disulfide transistors and circuits for high-speed flexible electronics. *Nat. Commun.* **5**, 5143 (2014).

49. Ross, J. S. *et al.* Electrically tunable excitonic light-emitting diodes based on monolayer WSe2 p–n junctions. *Nat. Nanotechnol.* **9**, 268–272 (2014).

50. Wang, Q. H., Kalantar-Zadeh, K., Kis, A., Coleman, J. N. & Strano, M. S. Electronics and optoelectronics of two-dimensional transition metal dichalcogenides. *Nature Nanotechnology* **7**, 699–712 (2012).

51. Rosker, M. J., Wise, F. W. & Tang, C. L. Femtosecond optical measurement of hot-carrier relaxation in GaAs, AlGaAs, and GaAs/AlGaAs multiple quantum well structures. *Appl. Phys. Lett.* **49**, 1726–1728 (1986).

52. Bar-Ad, S. & Bar-Joseph, I. Exciton spin dynamics in GaAs heterostructures. *Phys. Rev.*

*Lett.* **68**, 349–352 (1992).

53. Echenique, P. M. *et al.* Decay of electronic excitations at metal surfaces. *Surface Science Reports* **52**, 219–317 (2004).

54. Leo, K. *et al.* Coherent oscillations of a wave packet in a semiconductor double-quantum-well structure. *Phys. Rev. Lett.* **66**, 201–204 (1991).

55. Khitrova, G., Gibbs, H. M., Jahnke, F., Kira, M. & Koch, S. W. Nonlinear optics of normal-mode-coupling semiconductor microcavities. *Reviews of Modern Physics* **71**, 1591–1639 (1999).

56. Mai, C. *et al.* Many-body effects in valleytronics: Direct measurement of valley lifetimes in single-layer MoS2. *Nano Lett.* **14**, 202–206 (2014).

57. Nie, Z. *et al.* Ultrafast Electron and Hole Relaxation Pathways in Few-Layer MoS2. *J. Phys. Chem. C* **119**, 20698–20708 (2015).

58. Dal Conte, S. *et al.* Ultrafast valley relaxation dynamics in monolayer MoS2 probed by nonequilibrium optical techniques. *Phys. Rev. B - Condens. Matter Mater. Phys.* **92**, 235425 (2015).

59. Wallauer, R., Reimann, J., Armbrust, N., Güdde, J. & Höfer, U. Intervalley scattering in MoS2 imaged by two-photon photoemission with a high-harmonic probe. *Appl. Phys. Lett.* **109**, 162102 (2016).

60. Rivera, P. *et al.* Observation of long-lived interlayer excitons in monolayer MoSe2–WSe2 heterostructures. *Nat. Commun.* **6**, 6242 (2015).

61. Rohwer, T. *et al.* Collapse of long-range charge order tracked by time-resolved photoemission at high momenta. *Nature* **471**, 490–494 (2011).

62. Ulstrup, S. *et al.* Ultrafast Band Structure Control of a Two-Dimensional Heterostructure.

*ACS Nano* **10**, 6315–6322 (2016).

63. Grubišić Čabo, A. *et al.* Observation of Ultrafast Free Carrier Dynamics in Single Layer MoS2. *Nano Lett.* **15**, 5883–5887 (2015).

64. Brühwiler, P. A., Karis, O. & Mårtensson, N. Charge-transfer dynamics studied using resonant core spectroscopies. *Rev. Mod. Phys.* **74**, 703–740 (2002).

65. Menzel, D. Ultrafast charge transfer at surfaces accessed by core electron spectroscopies. *Chem. Soc. Rev.* **37**, 2212 (2008).

66. Eads, C. N., Bandak, D., Neupane, M. R., Nordlund, D. & Monti, O. L. A. Anisotropic attosecond charge carrier dynamics and layer decoupling in quasi-2D layered SnS2. *Nat. Commun.* **8**, 1369 (2017).

67. Racke, D. A. *et al.* Disrupted Attosecond Charge Carrier Delocalization at a Hybrid Organic/Inorganic Semiconductor Interface. *J. Phys. Chem. Lett.* **6**, 1935–1941 (2015).

68. Kelly, L. L. *et al.* Hybridization-Induced Carrier Localization at the C60/ZnO Interface. *Adv. Mater.* **28**, 3960–3965 (2016).

69. Jiang, H. Electronic Band Structures of Molybdenum and Tungsten Dichalcogenides by the *GW* Approach. *J. Phys. Chem. C* **116**, 7664–7671 (2012).

70. Zhang, P. *et al.* A precise method for visualizing dispersive features in image plots. *Rev. Sci. Instrum.* **82**, 043712 (2011).

71. Zhang, L. & Zunger, A. Evolution of Electronic Structure as a Function of Layer Thickness in Group-VIB Transition Metal Dichalcogenides: Emergence of Localization Prototypes. *Nano Lett.* **15**, 949–957 (2015).

72. Holub-Krappe, E. *et al.* Photoemission and EXAFS Study of Na on 2H-TaS2. in *Application of Particle and Laser Beams in Materials Technology* 653–660 (Springer

Netherlands, 1995). doi:10.1007/978-94-015-8459-3_46

73. Pettenkofer, C. *et al.* Cs deposition on layered 2H TaSe2 (0 0 0 1) surfaces: Adsorption or intercalation? *Solid State Commun.* **84**, 921–926 (1992).

74. Komesu, T. *et al.* Occupied and unoccupied electronic structure of Na doped MoS2(0001). *Appl. Phys. Lett.* **105**, 241602 (2014).

75. Zhao, W. & Ding, F. Energetics and kinetics of phase transition between a 2H and a 1T MoS$_2$ monolayer—a theoretical study. *Nanoscale* **9**, 2301–2309 (2017).

76. Kim, B. S., Rhim, J.-W., Kim, B., Kim, C. & Park, S. R. Determination of the band parameters of bulk 2H-MX2 (M = Mo, W; X = S, Se) by angle-resolved photoemission spectroscopy. *Sci. Rep.* **6**, 36389 (2016).

77. Li, D. *et al.* Polarized X-ray absorption spectra and electronic structure of molybdenite (2H-MoS2). *Phys. Chem. Miner.* **22**, 123–128 (1995).

78. Mattheiss, L. F. Band Structures of Transition-Metal-Dichalcogenide Layer Compounds. *Phys. Rev. B* **8**, 3719–3740 (1973).

79. Föhlisch, A. *et al.* Direct observation of electron dynamics in the attosecond domain. *Nature* **436**, 373–376 (2005).

80. Mårtensson, N. & Nyholm, R. Electron spectroscopic determinations of M and N core-hole lifetimes for the elements Nb—Te (Z = 41 − 52). *Phys. Rev. B* **24**, 7121–7134 (1981).

81. Garcia-Basabe, Y. *et al.* Ultrafast charge transfer dynamics pathways in two-dimensional MoS2 –graphene heterostructures: a core-hole clock approach. *Phys. Chem. Chem. Phys.* **19**, 29954–29962 (2017).

82. Margaritondo, G., Rowe, J. E. & Christman, S. B. Photoionization cross section of $d$-core levels in solids: A synchrotron radiation study of the spin-orbit branching ratio. *Phys.*

*Rev. B* **19**, 2850–2855 (1979).

83. Föhlisch, A. *et al.* Verification of the core-hole-clock method using two different time references: Attosecond charge transfer in c(4 × 2)S/Ru(0 0 0 1). *Chem. Phys. Lett.* **434**, 214–217 (2007).

84. Gao, P., Wang, L., Zhang, Y., Huang, Y. & Liu, K. Atomic-Scale Probing of the Dynamics of Sodium Transport and Intercalation-Induced Phase Transformations in $MoS_2$. *ACS Nano* **9**, 11296–11301 (2015).

85. Leng, K. *et al.* Phase Restructuring in Transition Metal Dichalcogenides for Highly Stable Energy Storage. *ACS Nano* **10**, 9208–9215 (2016).

86. Österlund, L., Chakarov, D. V. & Kasemo, B. Potassium adsorption on graphite(0001). *Surf. Sci.* **420**, 174–189 (1999).

87. Fischer, N., Schuppler, S., Fauster, T. & Steinmann, W. Coverage-dependent electronic structure of Na on Cu(111). *Surf. Sci.* **314**, 89–96 (1994).

88. Lindgren, S. A. & Walldén, L. Interband transitions, surface-barrier photoabsorption, and charge-screening waves observed via optical second-harmonic generation of Cs-covered Cu(111). *Phys. Rev. B* **45**, 6345–6347 (1992).

Supplementary Information

# Ultrafast Carrier Dynamics in Two-Dimensional Electron Gas-Like K-doped MoS$_2$

*Calley N. Eads,[1] Sara L. Zachritz,[1] Joohyung Park,[1] Anubhab Chakraborty,[1] Dennis L. Nordlund,[2] and Oliver L.A. Monti[1,3*]*

[1] University of Arizona, Department of Chemistry and Biochemistry, 1306 E. University Blvd., Tucson, AZ 85721, USA

[2] SLAC National Accelerator Laboratory, Stanford Synchrotron Radiation Lightsource, 2575 Sand Hill Rd., MS 99, Menlo Park, CA 94025, USA

[3] University of Arizona, Department of Physics, 1118 E. 4$^{th}$ St., Tucson, AZ 85721, USA

**\*Corresponding author:** Oliver L.A. Monti, monti@u.arizona.edu

**Supplementary Note 1. Potassium Intercalation**

To understand the fate of the K adsorbate, we performed surface-sensitive X-ray photoemission spectroscopy (XPS) of 0.2 ML K on $MoS_2$ at a photon energy of 600 eV and at grazing incidence angle. Supplementary Fig. 1 shows the evolution of XP feature K 2p over a 16 hr period. All XP spectra were normalized to the C 1s peak resulting from a small amount of adventitious carbon on the $MoS_2$ crystal surface. In the chosen combination of photon energy and incidence angle, XPS is exquisitely surface sensitive with a measurement depth of a no more than one $MoS_2$ layer. The XP spectra demonstrate therefore unequivocally that K atoms do not stay as surface adsorbates but rather intercalate and diffuse irreversibly into the crystal to a depth of at least one layer. There are no further changes to the XP or resonant photoemission spectra over the course of the measurement period, indicating that initial intercalation is fast while diffusion throughout the crystal is slow. This is also supported by the absence of any K 2p signatures after further cleaves of the $MoS_2$ crystal.

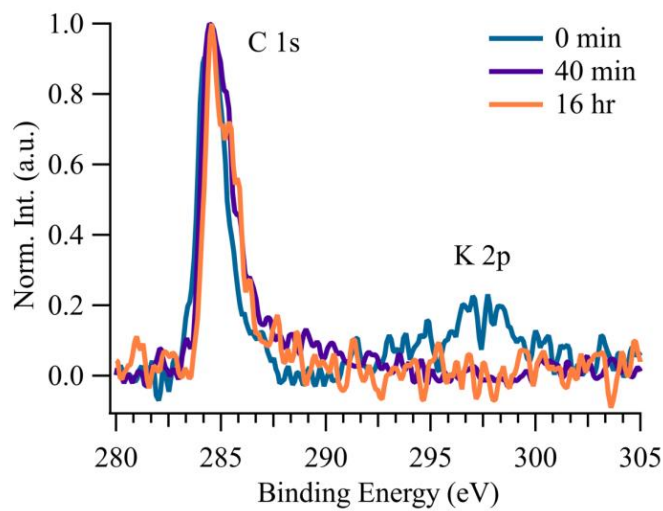

Supplementary Figure 1. XP spectra of C 1s and K 2p (normalized to C 1s) after K addition to bulk $MoS_2$ (0 min) as a function of time, which demonstrates prompt K diffusion into the substrate.

**Supplementary Note 2. XA Spectra of MoS$_2$ and K/MoS$_2$**

Though the Mo M$_3$ edge XA spectra exhibits a small shift and a drop in intensity, the overall shape remains the same. This is shown in Supp. Fig. 2 where both spectra are superimposed and fit with a Voigt profile to give nearly identical linewidths.

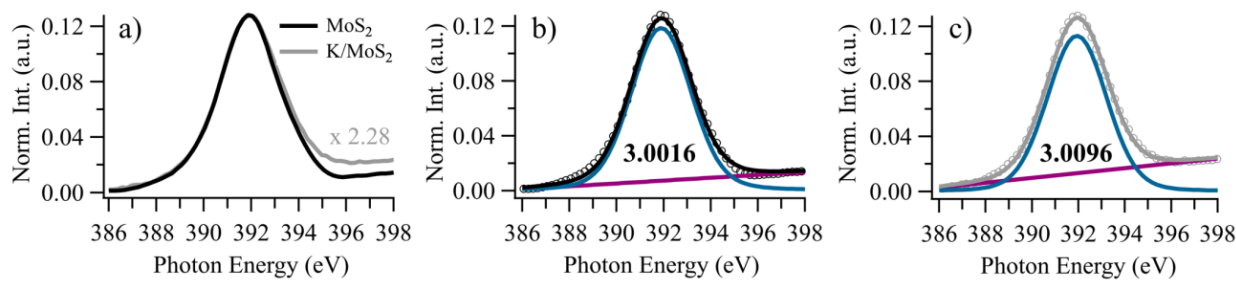

Supplementary Figure 2. XAS comparison of as-cleaved and K-dosed MoS$_2$ with associated fits. a) Comparison of XA spectra on the Mo M$_3$ edge for MoS$_2$ and K/MoS$_2$. The latter is scaled and shifted for ease of comparison. XA spectral fits using a linear background and Voigt peak profile for b) MoS$_2$ and c) K/MoS$_2$. Nearly identical FWHM values (in eV) are identified in parts b) and c).

**Supplementary Note 3. Core-Hole Clock Spectroscopy Scheme of Competitive Kinetics**

Core-hole clock spectroscopy (CHC) may be readily interpreted in the framework of a two-level scheme (ground state $|0\rangle$ and core-excited state $|1\rangle$) coupled to a continuum of states $|k\rangle$ composed of electronic states with crystal momentum $k$ (Supp. Fig. 3). Assuming that all decay channels are independent and that coherences can be neglected due to decay into dense continua, the decay of the core-excited state $|1\rangle$ may occur through a number of Auger decay processes with a combined rate $k_{c-h}$, or by delocalization into the continuum of electronic states $\{|k\rangle\}$ in the solid and characterized by rate $k_{CB}$. If $k_{CB} \ll k_{c-h}$, resonant excitation of $|1\rangle$ enhances the intensity of the photoemission features associated with the Auger-like decays, while such enhancement is suppressed if instead $k_{CB} \gg k_{c-h}$. The competition of these two decay processes, governed by the

relative magnitude of the two rate constants, determines therefore the intensity of the associated spectroscopic features in the resonant photoemission spectra.

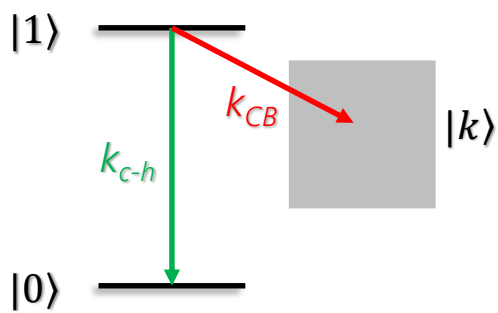

Supplementary Figure 3. Principle of CHC to determine lifetimes. The relative magnitude of the two rate constants $k_{c-h}$ and $k_{CB}$ determines the intensity of associated Auger features in the resonant photoemission map.